\documentclass[12pt]{article}
\usepackage{graphics}
\usepackage{amsmath}
\usepackage{amsfonts}
\usepackage{amssymb}
\begin{document}

\begin{center}
\Large {\bf Covariant Equilibrium Statistical Mechanics} \\ 
\vspace*{6mm}
\normalsize
\large E. Lehmann\footnote{ Electronic mail: Lehmann.Ewald@t-online.de}\\
\normalsize
Tannenweg 19\\
D-75394 Oberreichenbach\\
Germany\\
\end{center}

\normalsize
\begin{center}
\vspace*{5mm}
\today \\ \vspace*{5mm}
\end{center}
\vspace*{7mm}
\normalsize
~\\
%%%%%%%%%%%%%%%%%%%%%%%%%%%%%%%%%%%%%%%%%%%%%%%%%%%%%%%%%%%%%%%%%%%%%%%%%%%
%%%%%%%%%%%%%%%%%%%%%%%%%%%%%%%%%%%%%%%%%%%%%%%%%%%%%%%%%%%%%%%%%%%%%%%%%%%
{\large {\bf Abstract:}}
\\
\\
A manifest covariant equilibrium statistical mechanics is constructed starting
with a $8N$ dimensional extended phase space which is reduced to the $6N$ physical degrees of freedom using the
Poincar\'e-invariant constrained Hamiltonian dynamics describing the
micro-dynamics of the system.
The reduction of the extended phase space is initiated
forcing the particles on energy shell and fixing their individual time
coordinates with help of invariant time constraints.

The Liouville equation and the equilibrium condition are formulated in respect
to the scalar global evolution parameter which is introduced by the time
fixation conditions.
The applicability of the developed approach is shown for both, the perfect gas as well as the real gas.

As a simple application the canonical partition integral of the monatomic perfect gas is calculated and compared with other approaches.
Furthermore, thermodynamical quantities are derived.

All considerations are shrinked on the classical Boltzmann gas composed of massive particles and hence quantum effects are discarded.
\newpage
\section{Introduction and Motivation}

The motivation to investigate relativistic equilibrium statistical mechanics
is given by theoretical reasons as well as by practical reasons.
From the theoretical point of view equilibrium statistical mechanics must be
brought into accordance with the principles of relativity like all other branches of physics.
From the practical point of view there might be important relativistic effects at
(very) high temperatures.
Those effects may play a major role within cosmological problems.
Hence, equilibrium statistical mechanics is also interesting in the framework of
general relativity and as a proper base a manifest covariant approach of a
special relativistic equilibrium statistical mechanics is demanded.

While the covariant generalization of kinetic equations as a major application
of non-equilibrium statistical mechanics can be studied since many years with help
of textbooks (see e.g. \cite{GLW}) the relativistic equilibrium statistical mechanics
has not reached such stable states. (The interest on covariant kinetic equations
was mainly driven by practical reasons, i.e. by applications in various fields like
plasma physics, astrophysics, heavy ion reactions etc.).

Although J\"uttner presented already 1911 relativistic calculations concerning the perfect gas \cite{Juettner11}\footnote{Note, J\"uttners calculations are relativistic but not covariant.}
the manifest covariant description of equilibrium statistical mechanics is still under
consideration and a lot of attempts have been performed already to tackle the related problems.
An overview about such problems may be studied in older papers and reviews like the ones of
Hakim \cite{Hakim} and the one of Havas \cite{Havas} as well as references therein.

In this paper we mainly focus on the problem that the phase space $\int d\Gamma$ must be
conceived in an invariant fashion demanded e.g. by the invariance of the entropy
(i.e. $S=k$ln$\Gamma$).
Ter Haar and Wergeland \cite{HW71} have summarized very clear the two possibilities
which intuitively offer themselves:\\
"1) Either one makes $\Gamma$ explicitly invariant extending it from $6N$ to $8N$ dimensions.
That implies a many-time formulation of the underlying dynamics. For equilibrium theory, at any rate, this does not seem to be a suitable approach.\\
2) Or, one retains (in the classical limit case) the usual phase space and considers the Lorentz transformations as canonical. This will certainly accomplish the desired invariance, but requires certain auxiliary re-interpretations." \\
The work of Balescu et al. \cite{Balescu67} strongly enforces possibility 2) but applying the method
on non-equilibrium statistical mechanics only.\\
The refusing statement of ter Haar and Wergeland concerning possibility 1) was certainly
motivated by the fact that any straight forward attempt to formulate classical dynamics in a covariant fashion runs automatically into conflicts defined by the famous no-interaction theorem \cite{NIT}. This theorem states that only a collection of non-interacting particles can be described within a Poincare invariant fashion if a straight forward generalization of the usual dynamical description is applied.
But the conflicts with the no-interaction theorem can be avoided
within the framework of constrained Hamiltonian dynamics.
This formalism is based on an algorithm which was introduced by Dirac \cite{Dirac}
and Bergmann \cite{Bergmann55} in order to express a
theory based on a singular Lagrangian in a generalized phase space approach
with help of constraints\footnote{A general and introductional overview about this formalism can be found in monographs and textbooks (see \cite{CHD}).}.
In this context the original $8N$ dimensional extended phase space given by
$N$ interacting relativistic particles can be reduced to $6N$ dimensions with the help
of $2N$ constraints fixing the energies and the time coordinates of these $N$ particles.
Doing this a $6N$ dimensional hypersurface in the original $8N$
dimensional phase space is defined on which the system is allowed to move during its evolution.
The evolution of the system is described in respect to a global scalar
evolution parameter which is introduced by the time constraints relating
the individual time coordinates to this parameter.
This formalism is not changing the notion of simultaneity by a change of
the frame of reference, means one gets an invariant notion of simultaneity
as well as invariant world lines.
This method has been successfully applied
already in complicated non-equilibrium situations namely, in order to calculate heavy ion reactions
at intermediate and high energies (\cite{Sorge89},\cite{Lehmann}).

Hence we do not agree that possibility 1) is not a suitable way and we propose in this paper
an approach which uses the formalism of constrained Hamiltonian dynamics to describe the micro-dynamics of the system and which develops on this base an equilibrium statistical mechanics
in a manifest covariant fashion.

A similar approach has been presented already by Hsu \cite{Hsu83}.
But this approach is based on the so-called 'common relativity' (developed by the same author)
which is a rather special formulation and interpretation of relativity.
In contradiction to the work of Hsu we prefer in this paper to apply the usual framework and interpretion of relativity.

Another interesting approach following possibility 1) is the one of Horwitz et al. \cite{Horwitz81}
which is based on the many-body theory of the relativistic mechanics proposed by
St\"uckelberg \cite{Stueckelberg41}.
The main characteristics of this method are the
separation of the center of mass motion in the same way as in the non-relativistic theory
and the allowance of the particles being off-shell resulting finally in a new potential
corresponding to the mass degree of freedom of the relativistic system.
However, in our opinion, the extended phase space is not reduced in a satisfying way within this
approach: phase space integrations are only performed keeping a common time interval $\Delta t$
because no specific time fixations have been incorporated in this model
(see e.g. eq. (3.25) in \cite{Horwitz81}).

Furthermore, we would like to mention the paper of Tretyak \cite{Tretyak98}. This author
used the front form of dynamics to investigate relativistic equilibrium statistical mechanics.
But shrinking the phase space ad hoc to $6N$ dimensions this approach, although very elegant and usefull, can not be regarded to be manifest covariant.

The presentation of our approach in this paper is organized as follows: \\
In the following section 2 we give a brief overview on the constrained Hamiltonian dynamics in a representation which is suitable for our purpose. In section 3 we propose the formulation of
a covariant equilibrium statistical mechanics using the framework explained in section 2.
In section 4 we apply the developed approach to the canonical ensemble of the monatomic perfect gas. The obtained results are compared with other approaches. Finally, section 5 contains some concluding remarks.

Some remarks on the used notation are at order:\\
Throughout the paper relativistic units are used, i.e.
$\hbar\!=ß\!c\!=\!1$. The diagonal metric tensor has been chosen to be $(1,-1,-1,-1)$.
The usual summation convention (i.e. summation of repeated indices in the same expression) is applied in case of Greek indices (representing tensor indices)
which are running from 0 to 3. Latin indices indicate the particle index.
Capital letters are used for quantities characterizing the complete system while lower case letters
are used for quantities related to properties of single particles.
The complete set of $8N$ phase space coordinates will be abbreviated by
$(q^{\mu},p^{\mu})=(q_1^{\mu},...,q_N^{\mu},p_1^{\mu},...,p_N^{\mu})$.
A contravariant 4-vector $ a^{\mu} $ is defined as $ a^{\mu} = (a^0, {\it{\bf a}})$.
For the Poisson bracket of two dynamical phase space variables $A(q^{\mu},p^{\mu})$ and $B(q^{\mu},p^{\mu})$ the convention
\begin{equation}
\left\lbrace A,B\right\rbrace =\sum_i \left(
\frac{\partial A}{\partial p_i^{\mu}}\frac{\partial B}{\partial q_{i\mu}}
-\frac{\partial A}{\partial q_i^{\mu}}\frac{\partial B}{\partial p_{i\mu}}\right)
\end{equation}
is used.
As common within constrained Hamiltonian dynamics we also distinguish between weak equations and strong equations, assigned by '$\approx$' and '$=$', respectively.
While strong equations hold throughout the extended phase space weak
equations are demanded to be valid on the submanifold defined by the constraints.
Note, weak equations should not be used before all interesting Poisson brackets have been worked out.
\\
%%%%%%%%%%%%%%%%%%%%%%%%%%%%%%%%%%%%%%%%%%%%%%%%%%%%%%%%%%%%%%%%%%%%
%%%%%%%%%%%%%%%%%%%%%%%%%%%%%%%%%%%%%%%%%%%%%%%%%%%%%%%%%%%%%%%%%%%%
\section{Constrained Hamiltonian Dynamics}

The non-relativistic equilibrium statistical mechanics contains as an essential part
the description of the micro-dynamics using Hamiltonian dynamics, i.e. by classical propagation of particles under mutual interactions ''at-a-distance''.
The generalization of this non-relativistic particle dynamics to a
manifest covariant particle dynamics is not trivial because,
one has to know how to treat action-at-a-distance in a covariant
fashion. The conceptual problems in this field are formulated in
the no interaction theorem \cite{NIT}, which states that only a
collection of free particles can fulfill the following requirements:
\begin{itemize}
\item The many particle system is described in a Hamiltonian
formulation with a canonical representation of the Poincar\'e group.
\item The world lines are invariant (World line condition).
\item The physical coordinates are identified with the canonical coordinates.
\end{itemize}
At least one of these requirements must be violated if one is going to allow an interaction between the particles which depends only
on the trajectories of the particles.

In the seventies of the last century three possibilities have been worked out to avoid this negative statement
of the no-interaction theorem\footnote{An overview about the various attempts are collected in \cite{KL}.}:
The predictive relativistic mechanics, the singular Lagrange formalism and the constrained Hamiltonian dynamics.
All these formalisms have common that constraints are used to
reduce the original $8N$-dimensional phase space, which is given
by the space and time coordinates, the momenta and the energies
of the $N$ particles involved. The advantage of the constrained
Hamiltonian dynamics is
that one can choose these constraints more freely as in the
other methods. Furthermore, one can define $2N$ constraints in a
way obtaining the usual $6N$ dimensional phase space in the
non-relativistic limit, containing the physical dynamical
degrees of freedom. The constrained Hamiltonian dynamics
for particles under interaction was developed
by Bergmann, Anderson, Goldberg\cite{Bergmann55}, Todorov \cite{Todorov76}, Rohrlich \cite{Rohrlich79}, Komar
\cite{Komar78}, Sudarshan, Mokunda, Goldberg
\cite{Sudarshan81} and Samuel \cite{Samuel82}.
These authors picked up the
idea of Dirac \cite{Dirac} who has realized for the first time that
constraints are not only reducing the degrees of freedom but can
also determine the dynamics.

The model of Samuel contains as an important feature the cluster
decomposition in a satisfying way for our purpose. The cluster
decomposition, also called cluster separability, makes sure that when a
system of interacting particles breaks into dynamical
independent clusters, then the set of constraints has to break
into these dynamical independent clusters as well. The
disadvantage of the model of Samuel is that it is not
quantizable.
But this is playing no role for our purpose because, shrinking the
considerations on the classical Boltzmann gas, we need only
a classical Poincar\'e invariant description of particle dynamics
in the framework of an action-at-a-distance theory which
respects the cluster separability and the principle of causality.

In the following we demonstrate the concrete usage of this formalism to describe the
micro-dynamics of the perfect gas as well as of the real gas composed of massive particles.
\\
%%%%%%%%%%%%%%%%%%%%%%%%%%%%%%%%%%%%%%%%%%%%%%%%%%%%%%%%%%%%%%%%%%%%%%
\subsection{Micro-Dynamics of the Perfect Gas}

Modeling the micro-dynamics of the perfect relativistic gas a system of $ N $
non-interacting particles can be considered. The singular Lagrangian of such systems leads
directly to $ N $ on-shell constraints (see e.g. \cite{Barut64})
\begin{equation}
\varphi_i = \frac{1}{2m_i}\left( p_i^2-m_i^2\right)
\approx 0 \qquad ; \qquad i = 1,...,N,
\label{2.1}
\end{equation}
with $p_i^{\mu}$ being the 4-momentum of particle $i$ and $m_i$ being its rest mass.
In contradiction to an earlier work \cite{Lehmann} and the common usage
\cite{CHD} the mass-shell constraints have been modified by the scalar
factor $1/2m_i$. This modification provides several advantages:\\
First, the transition to the non-relativistic limit is simplified and
needs no further manipulations.\\
Second, this factor regulates the dimensions of the mass-shell constraints to be the one of an energy instead of an energy squared.\\
Third, and this is our major motivation, a straight forward performance
of phase space integrals is guaranteed by this modification.\\ 
Since these primary constraints are first class,
i.e. $ \left\lbrace \varphi_i,\varphi_j \right\rbrace \approx 0$,
no secondary constraints exist in this case. Nevertheless, the $ \varphi_i $ have to be
supplemented by $ N $ time fixation conditions which gauge the individual time coordinates
parameterizing the world lines.
Because non-interacting particles must not be correlated one often uses the simple gauging conditions
\begin{equation}
\chi_i = q_i^0-\tau \approx 0 \qquad ; \qquad i = 1,...,N,
\label{2.2}
\end{equation} 
which force the time coordinates of the particles to be equal to the global scalar evolution parameter $\tau$ in any inertial system.
These constraints are obviously not covariant.\\
But our aim in this paper is a manifest covariant reduction of the extended phase space and hence the constraints (\ref{2.2}) are not satisfying in order to treat the relativistic perfect gas.
Therefore, we define here another set of time fixations
\begin{equation}
\chi_i = \frac{p_i^{\mu}q_{i\mu}}{m_i}-\tau \approx 0
\qquad ; \qquad i = 1,...,N ,
\label{2.3}
\end{equation} 
which are covariant and also do not create any correlation among the particles
under consideration.
The acting of these gauging conditions is easy recognized when considering it in
the rest frame of a particle $ i $: In this particular frame of reference 
$ p_i^{\mu}=(m_i, {\it {\bf 0}}) $
and $ \chi_i $ reduces to
\begin{equation}
\chi_i = q_{i0}-\tau \approx 0,
\label{2.4}
\end{equation}
i.e. the time coordinates of the particles equal the global evolution parameter $ \tau $ in their rest frame.
In other words, using (\ref{2.3}) the proper times of the particles are synchronized to $ \tau $
within this covariant treatment.
Note, both time fixations, (\ref{2.2}) and (\ref{2.3}) are first class, i.e.
$ \left\lbrace \chi_i,\chi_j \right\rbrace \approx 0$,
but the complete set of $ 2N $ constraints is second class since
$ \left\lbrace \varphi_i,\chi_j \right\rbrace \not\approx 0$.

Because the canonical Hamiltonian derived by a Legendre transformation from the singular Lagrangian
of a system of relativistic particles is identical vanishing \cite{Barut64} the Dirac Hamiltonian can be constructed by a linear combination of the $ \tau $-independent constraints:
\begin{equation}
H = \sum_{i=1}^{N}\lambda_i\varphi_i.
\end{equation} 
This Hamiltonian generates the equations of motion in respect to $ \tau $ to be
\begin{equation}
\frac{dq_i^{\mu}}{d\tau} = \left\lbrace H,q_i^{\mu}\right\rbrace \approx \sum_{j=1}^N \lambda_j
\frac{\partial \varphi_j}{\partial p_{i\mu}},
\label{2.5}
\end{equation}
\begin{equation}
\frac{dp_i^{\mu}}{d\tau} = \left\lbrace H,p_i^{\mu}\right\rbrace \approx - \sum_{j=1}^N \lambda_j
\frac{\partial \varphi_j}{\partial q_{i\mu}}.
\label{2.6}
\end{equation}
The Lagrange multipliers $ \lambda_i $ can be determined with help of the consistency
conditions
\begin{equation}
\frac{d\psi_j}{d\tau} 
= \frac{\partial\psi_j}{\partial\tau} + \left\lbrace H,\psi_j\right\rbrace
\approx \frac{\partial\psi_j}{\partial\tau} + 
  \sum_{i=1}^N \lambda_i\left\lbrace \varphi_i,\psi_j\right\rbrace
\approx 0 \quad ; \quad j=1,...,2N ; 
\label{2.7}
\end{equation}
which guarantee the validity of the constraints during the complete
evolution of the system and which have to be valid for the complete
set of $ 2N $ constraints
\begin{equation}
\label{2.8}
\psi_j =
  \begin{cases}
    \varphi_j & ; 0 < j \leq N ,\\
    \chi_{j-N}& ; N < j \leq 2N.
  \end{cases}
\end{equation}
Introducing the invertible $N \times N$-matrix $C$ with elements
$ C_{ij} =\left\lbrace \varphi_i, \chi_j \right\rbrace  $
the Lagrange multipliers are uniquely determined by
\begin{equation}
\lambda_i \approx -\sum_{j=1}^{N}\frac{\partial \chi_j}{\partial \tau} C_{ij}^{-1}
\label{2.9}
\end{equation} 
to be $\lambda_i = m_i/p_i^0$ if using (\ref{2.2}) and $\lambda_i = m_i^2/p_i^2$ if using (\ref{2.3}).

In the following we will use the term {\bf \textit{semi-covariant approach}} whenever
the set of $ 2N $ constraints is assumed to consist of the $ N $ on shell constraints
(\ref{2.1}) and the simple time fixations (\ref{2.2}) because $ N $ covariant constraints
are supplemented by $ N $ non-covariant constraints.
Furthermore, we will use the term {\bf \textit{full covariant approach}} whenever the on-shell
constraints (\ref{2.1}) are combined with the covariant time fixations (\ref{2.3}) to
build the complete set of $ 2N $ constraints.

Using $ p_i^0=\varepsilon_i+m_i $ the on-shell constraints
reduce in the non-relativistic limit ($ \varepsilon_i \ll m_i $) to
\begin{equation}
\varphi_i = \varepsilon_i - \frac{{\it {\bf p}}_i^2}{2m_i} \approx 0.
\label{2.11} 
\end{equation}
Furthermore, the covariant time fixations (\ref{2.3}) reduce to the simple time fixations (\ref{2.2}) in the non-relativistic limit
and the global evolution parameter $ \tau $ is identified
with the absolute time $ t $.
Consequently, the usual non-relativistic Hamiltonian dynamics is obtained as generated by the non-relativistic Hamiltonian
\begin{equation}
H^{(nr)} = \sum_{i=1}^{N}\frac{{\it {\bf p}}_i^2}{2m_i}.
\label{2.12}
\end{equation}
\\
%%%%%%%%%%%%%%%%%%%%%%%%%%%%%%%%%%%%%%%%%%%%%%%%%%%%%%%%%%%%%%%%%%%%%%
\subsection{Micro-Dynamics of the Real Gas}

In contradiction to ideal systems like the perfect gas real systems can only be understood
if the intermolecular interactions are taken into account when describing the related
micro-dynamics.
Applying the formalism of the constrained Hamiltonian dynamics to the micro-dynamics
of the real gas we therefore have to consider a system of $ N $ particles in mutual interaction.
In this case we prefer to use a set of constraints quite similar to the constraints which have been successfully applied already within non-equilibrium situations when simulating heavy ion reactions
(see \cite{Sorge89},\cite{Lehmann}).
The first $N$ constraints are chosen as on-shell constraints
\begin{equation}
\varphi_i = \frac{1}{2m_i}\left(p_i^2 - m_i^2 -\tilde V_i\right)  \approx 0 \qquad ; \qquad i=1,...,N .
\label{2.20}
\end{equation}
Note, as already discussed in section 2.1, the on-shell constraints
have been modified by the scalar factor $1/2m_i$.
Regarding the potential part $\tilde V_i$
this choice of the constraints requires that
$\tilde V_i$ should be a Lorentz scalar and therefore
a function of Lorentz scalars. Defining a
system with mutual two-body interactions, $\tilde V_i$ should be given by a sum of
these two-body interactions
\begin{equation}
\tilde V_i = \sum_{j\neq i}\tilde V_{ij}(q^2_{Tij}).
\label{2.21}
\end{equation}
Using equation (\ref{2.21}) we assume that the two-body interactions depend only on the
Lorentz invariant squared transverse distance
\begin{equation}
q^2_{Tij} = q^2_{ij} - \frac{(q^{\mu}_{ij}p_{ij\mu})^2}{p_{ij}^{\,\,2}} ,
\label{2.22}
\end{equation}
with $q^{\mu}_{ij}=q^{\mu}_i-q^{\mu}_j$ being the four
dimensional distance and $p^{\mu}_{ij}=p^{\mu}_i+p^{\mu}_j$ the
sum of the momenta of the two interacting particles $i$ and $j$.
This particular choice has two advantages:\\
First, in the common CMS of two interacting particles CMS$_{ij}$
$q^2_{Tij}$ reduces to $-{\it {\bf q}}_{ij}^2$ which is quite natural
since only in the
CMS$_{ij}$ particle $i$ and particle $j$ can be treated on the
same footing. In addition, in the non-relativistic limit the
potential part $\tilde V_i$ can be connected with the corresponding
potential $V_i$ used in the non-relativistic theory:
Using $ p_i^0=\varepsilon_i+m_i $ and regarding the non-relativistic limit ($ \varepsilon_i \ll m_i $) the on-shell constraints reduce to
\begin{equation}
\varphi_i = \varepsilon_i - \frac{{\it {\bf p}}_i^2}{2m_i}
- \frac{{\tilde V_i}}{2m_i} \approx 0.
\label{2.23}
\end{equation}
The comparison of (\ref{2.23}) with the usual expression of the
non-relativistic energy suggests
\begin{equation}
\tilde V_i = 2m_iV_i ,
\label{2.25}
\end{equation}
and the two-body potential acting among two particles $i$ and $j$
felt by particle $i$ reads
in the common CMS$_{ij}$
\begin{equation}
\tilde V_{ij}(-{\it {\bf q}}^2_{ij})=2m_i V_{ij}(-{\it {\bf q}}^2_{ij}) ,
\label{2.26}
\end{equation}
i.e. a non-relativistic two-body potential can be generalized to be used in the relativistic theory.
For instance, the widely used Lennard-Jones potential can be generalized replacing the squared
distance ${\it {\bf q}}^2_{ij}$ by the Lorentz invariant squared transverse distance $q^{2}_{Tij}$ obtaining
\begin{equation}
\tilde V_{ij}^{(LJ)}(q^2_{Tij})=2m_i \kappa
\left[\;\left( \frac{\sigma}{\sqrt{q^2_{Tij}}}\right)^{\!12} - \left( \frac{\sigma}{\sqrt{q^2_{Tij}}}\right)^{\!6}\;\right]
\end{equation}
with the two 'tuning parameters' $ \kappa $ and $ \sigma $ having the dimensions of an energy and
a length, respectively.
Note, due to the second term of the right
hand side of equation (\ref{2.22}) this generalized interaction
is slightly implicit momentum dependent. Because
of this term, which gives the longitudinal squared distance, the
interaction depends not only on the distance of the two
interacting particles but also on the direction of
their center of mass motion in the rest frame of the real gas.

Since the on-shell constraints alone do not specify the world
lines one needs additional $N$ constraints which are fixing the
relative times of the particles. Like in the model of Samuel \cite{Samuel82}
this second set of $N$ constraints should fulfill the following conditions:\\
1) $N-1$ of them should be Poincar\'e invariant in order to
fulfill the required world line invariance.\\
2) Causality has to be respected.\\
3) One has to be consistent with cluster separability.\\
4) A global evolution parameter must be defined.\\
These conditions can be fulfilled by defining the following set of $N$ time
fixations
\begin{equation}
\chi_{i} = \frac{1}{m_i}\sum_{j(\neq i)}\omega_{ij}p^{\mu}_{ij}q_{ij\mu} \approx 0 \qquad ;
\qquad i=1,...,(N-1)
\label{2.27} 
\end{equation}
\begin{equation}
\chi_{N} = P^{\mu}Q_{\mu}-\tau \approx 0
\label{2.28}
\end{equation}
with $P^{\mu}=P^{\mu}/\sqrt{P^2}$, $P^{\mu}=\sum_i p^{\mu}_i$,
$Q^{\mu}=\frac{1}{N}\sum_i q^{\mu}_i$ and the dimensionless scalar
weighting function
\begin{equation}
\omega_{ij} = \frac{1}{q^2_{ij}/\sigma^2}
\,\mbox{e}^{(q^2_{ij}/\sigma^2)}
\label{2.29}
\end{equation}
with $\sigma$ being a characteristic interaction distance of the modelled interaction. Note, the scalar factor $1/m_i$ regulates the dimension of the time fixations to be the one of a length.
The conditions (\ref{2.27}) are motivated by
studies in the framework of Singular Lagrangians \cite{Dominici78}.
Using this methods one gets up to the weighting functions
$\omega_{ij}$ the same conditions as secondary constraints and the
on-shell conditions as primary constraints in a natural way.
The important fact for using the expression (\ref{2.29}) as weighting
function is that this scalar function respects the principle of causality,
while the ones used in the Singular Lagrangian theories and in
the model of Samuel can violate this important physical restriction
(see \cite{Sorge89}).

The constraints (\ref{2.27}) take care that the times of interacting particles
are not dispersed to much in their common CMS$_{ij}$.
Furthermore, they specify the dynamics by fixing the times at
which the forces have to be calculated but they do not specify the
global evolution parameter $\tau$. This evolution parameter must
defined to be determined dynamically since the no-interaction theorem
can only be avoided in this way and is defined by the gauging condition
(\ref{2.28}) in a way, that the individual times are
increasing with increasing $\tau$.\\
In the CMS of all particles involved $\tau$ is simply given by
the average of all time coordinates
\begin{equation}
\tau = P^{\mu}Q_{\mu} \stackrel{\rm CMS}{\longrightarrow}
\frac{1}{N}\sum_{i=1}^{N}q_i^0 .
\label{2.30} 
\end{equation}

This set of $2N$ constraints given by (\ref{2.20}),(\ref{2.27}) and (\ref{2.28})
determines covariant world lines parameterized by the initial
data at equal $\tau$.

A linear combination of the $2N-1$ $\tau$-independent constraints
\begin{equation}
\label{2.31}  
\psi_i = 
  \begin{cases} \varphi_i & ; 0 < i \leq N \\
                \chi_{i-N} & ; N < i \leq 2N-1
  \end{cases}
\end{equation}
constructs the Dirac Hamiltonian
\begin{equation}
H = \sum_{i=1}^{2N-1}\lambda_i\psi_i
\label{2.32}
\end{equation}
which generates the equations of motion in respect to $ \tau $ to be
\begin{equation}
\frac{dq_i^{\mu}}{d\tau} \approx \sum_{j=1}^{2N-1} \lambda_j
\frac{\partial \psi_j}{\partial p_{i\mu}},
\label{2.33}
\end{equation}
\begin{equation}
\frac{dp_i^{\mu}}{d\tau} \approx - \sum_{j=1}^{2N-1} \lambda_j
\frac{\partial \psi_j}{\partial q_{i\mu}}.
\label{2.34}
\end{equation}
The Lagrange multipliers $ \lambda_i $ can be determined with help of the consistency conditions
\begin{equation}
\frac{d\psi_j}{d\tau}
= \frac{\partial\psi_j}{\partial\tau} + \left\lbrace H,\psi_j\right\rbrace 
\approx \frac{\partial\psi_j}{\partial\tau} +
  \sum_{i=1}^{2N-1} \lambda_i\left\lbrace \psi_i,\psi_j\right\rbrace 
\approx 0 \quad ; \quad j=1,...,2N ;
\label{2.7a}
\end{equation}
and are finally given by
\begin{equation}
\lambda_i \approx -\frac{\partial \chi_N}{\partial\tau} 
C^{-1}_{i\,2\!N}
\label{2.35}
\end{equation}
with $C$ being the invertible $2N \times 2N$-matrix with elements 
$ C_{ij}=\left\lbrace \psi_i,\psi_j\right\rbrace $.

The whole formalism has a well defined non-relativistic limit as has been shown by Sorge \cite{Sorge89}.
In leading order of $1/c$ one gets for the
first $2N-1$ constraints
\begin{equation}
\varphi_i = \varepsilon_i-\frac{{\it {\bf p}}_i^2}{2m_i} -\frac{\tilde V_i}{2m_i} \approx 0
\qquad ;
\qquad i=1,...,N
\label{2.36}
\end{equation}
\begin{equation}
\chi_i = q_i^0 - q_N^0 \approx 0 \qquad ; \qquad i=1,...,N-1
\label{2.37}
\end{equation}
which generate the same dynamics as the non-relativistic many
body Hamiltonian
\begin{equation}
H^{(nr)} = \sum_{i=1}^N \left(\frac{{\it {\bf p}}_i^2}{2m_i} + V_i \right) .
\label{2.38}
\end{equation}
\\
%%%%%%%%%%%%%%%%%%%%%%%%%%%%%%%%%%%%%%%%%%%%%%%%%%%%%%%%%%%%%%%%%%%%%%
%%%%%%%%%%%%%%%%%%%%%%%%%%%%%%%%%%%%%%%%%%%%%%%%%%%%%%%%%%%%%%%%%%%%%%
\section{Covariant Statistical Mechanics in the Framework of 
Constrained Hamiltonian Dynamics }

Regarding the constrained Hamiltonian dynamics as a proper formalism to
describe a system of interacting particles in a covariant fashion we are
to develop a manifest covariant statistical mechanics on its base.
We start this development considering the extended phase space.
In case of a system involving $N$
particles this invariant phase space has $8N$ dimensions given by the $N$ 4-momenta
$p_i^{\mu}$ and the $N$ 4-position-vectors $q_i^{\mu}$ ($i=1,..,N$).\\
The phase space distribution function $D(q^{\mu},p^{\mu},\tau)$
representing the statistical ensemble is assumed to depend on the complete set of $8N$
phase space coordinates and, in general, on the global evolution parameter $\tau$
which is introduced by certain time fixation conditions (compare section 2).
Consequently, the Liouville equation and the equilibrium conditions have to be
discussed in respect to $\tau$.

The Liouville equation is thus given by
\begin{equation}
\frac{d D(q^{\mu},p^{\mu},\tau)}{d\tau} = 0 \quad \Longleftrightarrow \quad
\frac{\partial D(q^{\mu},p^{\mu},\tau)}{\partial \tau} =
\left\lbrace D(q^{\mu},p^{\mu},\tau) , H \right\rbrace .
\label{3.1}
\end{equation}
The micro-dynamics of the system is generated by the Hamiltonian
\begin{equation}
H = \sum_{i=1}^M \lambda_i\psi_i 
\label{3.2}
\end{equation}
where $M\!=\!N$ in case of the perfect gas and $M\!=\!2N\!-\!1$ in case of the real gas using the constraints as introduced in section 2.\\
Hence we get
\begin{equation}
\frac{\partial D(q^{\mu},p^{\mu},\tau)}{\partial \tau}
\approx \sum_{i=1}^M 
\lambda_i \left\lbrace  D(q^{\mu},p^{\mu},\tau) , \psi_i \right\rbrace .
\label{3.3}
\end{equation}

The reduction of the invariant phase space volume element of the extended phase space
\begin{equation}
d \Gamma_e = \prod_{i=1}^N d^4\!q_i d^4\!p_i.
\label{3.4}
\end{equation}
is performed incorporating all $2N$ constraints by means of 
$\delta$-functions, i.e.
\begin{equation}
d \Gamma = \prod_{i=1}^N d^4\!q_i d^4\!p_i \,\Theta(p_i^0)
\delta(\varphi_i) \delta(\chi_i).
\label{3.5}
\end{equation}
The Heaviside-$\Theta$-function has been added in order to shrink on particles with
positive energy which is usefull if classical particles are considered only.

Equilibrium distribution functions are $\tau$ independent solutions of (\ref{3.1}) and hence have to fulfill the condition
\begin{equation}
\sum_{i=1}^M \lambda_i \left\lbrace D(q^{\mu},p^{\mu}) , \psi_i \right\rbrace  \approx 0.
\label{3.6}
\end{equation}
Using (\ref{3.5}) the ensemble average (i.e. the classical expectation value)
of a measurable quantity $A(p^{\mu},q^{\mu})$ at equilibrium is given by
\begin{equation}
<\!A\!> = \frac{
\int \prod_{i=1}^N d^4\!q_i d^4\!p_i \,\Theta(p_i^0) \delta(\varphi_i) \delta(\chi_i) A(p^{\mu},q^{\mu}) D(q^{\mu},p^{\mu})}
{\int \prod_{i=1}^N d^4\!q_i d^4\!p_i \,\Theta(p_i^0) \delta(\varphi_i) \delta(\chi_i)
D(q^{\mu},p^{\mu})}.
\label{3.7}
\end{equation}
We mention two simple (and widely used) possibilities to construct solutions of the Liouville equation fulfilling the equilibrium condition:
\begin{itemize}
\item
Constant phase space distributions $ D = $ const.
\item
Phase space distributions depending on quantities $ X_j $ which are conserved
during the evolution in respect to $ \tau $, i.e. $dX_j/d\tau=0$. \\
Assuming $ X_j \neq X_j(\tau) $ the condition
\begin{equation}
\sum_{i=1}^M \lambda_i \left\lbrace  X_j , \psi_i \right\rbrace
\approx 0.
\label{3.8}
\end{equation}
has to be valid in case of such conserved quantities.
\end{itemize}
Beside the trivial uniform ensemble the microcanonical ensemble can be regarded to be the most typical example
for the former case while the canonical as well as the grand canonical ensembles are
examples of the latter case.

In the non-relativistic theory the energy is playing a prominent role as
the most important conserved quantity.
In a relativistic generalization the energy has to be replaced by the 4-momentum
$ P^{\mu} $ or even by the energy-momentum tensor
$ T^{\mu\nu} $ of the system. A fruitfull alternate is the invariant energy
$E=U_{\mu}P^{\mu}$ with $U_{\mu}$ being the 4-velocity of the system. In
 the comoving frame of reference (i.e the restframe of the system) we get $E=U_{0}P^{0}=P^{0}$,
i.e. the invariant energy is nothing but the total (internal) energy of the system measured in its restframe.
We will use this invariant energy when generalizing the concepts of non-relativistic
statistical mechanics.

In the following we consider briefly the main equilibrium distributions:
\\
\\
%%%%%%%%%%%%%%%%%%%%%%%%%%%%%%%%%%%%%%%%%%%%%%%%%%%%%%%%%%%%%%%%%%%%%%
{\bf i) Microcanonical Ensemble}\\
\\
Considering the classical (i.e. non quantum mechanical) case only and applying our concept of the
invariant energy $E$ we can write the
distribution function of the microcanonical ensemble using the $\delta$-function
(compare e.g. \cite{Muenster}) as
\begin{equation}
D_M = \frac{1}{Z_M}\,\delta(E-E_0)
\label{3.20}
\end{equation}
with $E_0$ being the invariant energy of the closed system. The 'microcanonical partition
integral $Z_M$' (which is nothing but the phase space volume $\Gamma$) has to be
determined via
\begin{equation}
Z_M = \frac{1}{N!}\int\prod_{i=1}^N d^4\!q_i d^4\!p_i
\,\Theta(p_i^0) \delta(\varphi_i) \delta(\chi_i) \delta(E-E_0).
\label{3.21}
\end{equation}
The factor $1/N!$ respects the fact that the particles can not be distinguished. 
\\
\\
%%%%%%%%%%%%%%%%%%%%%%%%%%%%%%%%%%%%%%%%%%%%%%%%%%%%%%%%%%%%%%%%%%%%%%
{\bf ii) Canonical Ensemble}\\
\\
Considering the canonical ensemble the condition
\begin{equation}
\sum_{i=1}^M \lambda_i \left\lbrace  E , \psi_i \right\rbrace =
\sum_{i=1}^M \lambda_i \left\lbrace  U_{\mu}P^{\mu} , \psi_i \right\rbrace \approx 0
\label{3.9}
\end{equation}
is to be proved in order to use the ansatz
\begin{equation}
D_C(E) = \frac{1}{Z_C}\,\mbox{e}^{-\beta U_{\mu}P^{\mu}}
\label{3.10}
\end{equation}
for the canonical distribution function with the canonical partition integral \\
\begin{equation}
Z_C = \frac{1}{N!}\int\prod_{i=1}^N d^4\!q_i d^4\!p_i
\,\Theta(p_i^0) \delta(\varphi_i) \delta(\chi_i)\,\mbox{e}^{-\beta U_{\mu}P^{\mu}}.
\label{3.11}
\end{equation}
\\
\\
%%%%%%%%%%%%%%%%%%%%%%%%%%%%%%%%%%%%%%%%%%%%%%%%%%%%%%%%%%%%%%%%%%%%%%
{\bf iii) Grand Canonical Ensemble}\\
\\
The grand canonical ensemble is characterized by another conserved quantity beside the energy
namely, the particle number $N$.
Hence, generalizing the canonical ensemble accordingly leads to
\begin{equation}
D_G(E,N) = \frac{1}{Z_G}\,\mbox{e}^{-\beta U_{\mu}P^{\mu} -\alpha N}
\label{3.30}
\end{equation}
for the grand canonical distribution function with the grand canonical partition integral \\
\begin{equation}
Z_G = \frac{1}{N!}\int\prod_{i=1}^N d^4\!q_i d^4\!p_i
\,\Theta(p_i^0)  \delta(\varphi_i) \delta(\chi_i)
\,\mbox{e}^{-\beta U_{\mu} P^{\mu} -\alpha N}.
\label{3.31}
\end{equation}  
\\
\\
%%%%%%%%%%%%%%%%%%%%%%%%%%%%%%%%%%%%%%%%%%%%%%%%%%%%%%%%%%%%%%%%%%%%%%
{\bf iv) Related Thermodynamics}\\
\\
Let us briefly discuss the corresponding thermodynamics
generated by our approach. Like in the next section, we shrink on the canonical ensemble.
The formal properties of this thermodynamics are
dominated by the introduction of the invariant energy $E$.
As in the non-relativistic theory the free energy can be determined with help of the canonical partition integral $Z_C$ by
\begin{equation}
F = -kT \mbox{ln} Z_C.
\label{3.12}
\end{equation}
Like the invariant energy $E$ the free energy $F$ (as all other thermodynamical potentials) is defined to be a scalar. Furthermore, the temperature $T$ introduced by the
parameter $\beta=1/kT$ ($k$ being the Boltzmann constant) is a scalar quantity and the introduction of a heat vector
\cite{books} is superfluous\footnote{Of course, formally we get the same exponential factor (see eq. (\ref{3.10})) but the 4-velocity of the
system $U_{\mu}$ is now
defining the invariant energy $E=U_{\mu} P^{\mu}$ and not used to define the heat vector $\Theta_\mu=\frac{U_\mu}{T}$ like in other approaches \cite{books}. One can argue about this interpretation but the advantage defining an invariant energy is that this scalar quantity of the exponent can be demanded to be conserved during evolution in respect to $\tau$, i.e. to fulfill equation (\ref{3.8}).}.
Considering the usual relations defining other macroscopic state variables
like the entropy
\begin{equation}
S = k \mbox{ln} Z_C - \frac{1}{T}\frac{\partial \mbox{ln} Z_C}{\partial \beta}
\label{3.13}
\end{equation}
and the pressure
\begin{equation}
P = - \left( \frac{\partial F}{\partial V} \right)_T
\label{3.14}
\end{equation}
we stress that both, the macroscopic state variables as well as the
thermodynamical potentials are defined to be scalar quantities in our
approach. Hence, an artifical extension of those quantities to higher rank tensors as often done when formulating relativistic
thermodynamics (compare e.g. \cite{books}) is not needed.\\
The average energy is given by
\begin{equation}
\left\langle E\right\rangle 
= - \frac{\partial \mbox{ln}Z_C}{\partial \beta},
\label{3.15}
\end{equation}
and the specific heat can be determined with help of it:
\begin{equation}
c_V = \left( \frac{\partial \left\langle E\right\rangle }{\partial T}\right)_V.
\label{3.16}
\end{equation}

Looking to the grand canonical ensemble we note that the chemical potential $M$
introduced by the parameter $\alpha = \beta M = M/kT$ is a scalar as well in the approach considered.
\\
%%%%%%%%%%%%%%%%%%%%%%%%%%%%%%%%%%%%%%%%%%%%%%%%%%%%%%%%%%%%%%%%%%%%%
%%%%%%%%%%%%%%%%%%%%%%%%%%%%%%%%%%%%%%%%%%%%%%%%%%%%%%%%%%%%%%%%%%%%%
\section{The Canonical Ensemble of the Monatomic Perfect Gas}

In this section we want to apply the formalism developed in the previous sections.
Unfortunately, concrete calculations
(e.g. determining the canonical partition integral via equation (\ref{3.11}))
are note easy to be performed due to the structure of the phase space volume
element\footnote{Especially, as in the non-relativistic theory, calculations considering the real gas turn out to be complicated
depending on the underlying model of intermolecular interactions.}.
Hence, as an example, we just apply the formalism to the canonical ensemble of the
monatomic perfect gas.
In order to determine the canonical partition integral using formula (\ref{3.11})
we have to prove the validity of condition (\ref{3.9}):\\
By definition, the perfect gas is a system of non-interacting particles
and as a consequence the Hamiltonian contains the simple mass shell constraints (\ref{2.1}).
Hence, the partial derivatives of $  \varphi_i $ in respect
to $q_{j \mu}$ are vanishing and condition (\ref{3.9}) is fulfilled in case of the perfect gas.

Unfortunately, no analytical solution of the canonical partition integral
(\ref{3.11}) can be achieved within the full covariant approach.
Hence, numerical solutions have been worked out as will be explained
in section 4.2. \\
But before discussing these numerical results let us first present
another approach, namely the semi-covariant approach which provides
analytical results and hence can be easy compared with
other approaches known from the literature, e.g. the non-covariant
J\"uttner approach \cite{Juettner11}.
\\
%%%%%%%%%%%%%%%%%%%%%%%%%%%%%%%%%%%%%%%%%%%%%%%%%%%%%%%%%%%%%%%%%%%%%%
\subsection{Semi-covariant Approach}

In the semi-covariant approach the simple time fixation conditions (\ref{2.2})
are used and the partition integral reads
\begin{equation}
Z_C = \frac{1}{N!}\int \prod_{i=1}^N d^4\!q_i d^4\!p_i
\,\Theta(p_i^0) \delta\left( \frac{1}{2m_i}(p_i^2-m_i^2)\right) \delta(q_i^0-\tau)
\,\mbox{e}^{-\beta U_{\mu}P^{\mu}}.
\label{4.5}
\end{equation}
Using the properties
\begin{equation}
\label{4.d1}
\delta(ax) = \frac{1}{|a|}\delta(x)
\end{equation}
and
\begin{equation}
\label{4.d2}
\delta(x^2-a^2) = \frac{1}{2a}\left[ \delta(x-a)-\delta(x+a)\right]
\end{equation}
of the $\delta$-function we get
\begin{equation}
Z_C = \frac{1}{N!}\int \prod_{i=1}^N d^4\!q_i d^4\!p_i
\,\frac{m_i}{p_i^0}
\delta\left( p_i^0-\sqrt{{\it {\bf p}}_i^2+m_i^2}\right)
\delta\left( q_i^0 - \tau \right)\,\mbox{e}^{-\beta U_{\mu}P^{\mu}} .
\label{4.6}
\end{equation}
For the further concrete calculation we assume a comoving frame of reference
(i.e. the restframe of the gas). In this frame of reference we get
\begin{equation}
E = U_{\mu}P^{\mu} = U_{\mu} \sum_{i=1}^N p_i^{\mu} = \sum_{i=1}^N p_i^0
\label{4.7}
\end{equation}
and integrating the 0-components gives according the $\delta$-functions
\begin{eqnarray}
Z_C
& = &
\frac{1}{N!} \int \prod_{i=1}^N d^3\!q_i d^3\!p_i\,
\frac{m_i}{\sqrt{{\it {\bf p}}_i^2 + m_i^2}}
\,\mbox{e}^{-\beta \sqrt{{\it {\bf p}}_i^2 + m_i^2}} \nonumber \\
& = &
\frac{V^N}{N!} \int \prod_{i=1}^N d^3\!p_i\,
\frac{m_i}{\sqrt{{\it {\bf p}}_i^2 + m_i^2}}
\,\mbox{e}^{-\beta \sqrt{{\it {\bf p}}_i^2 + m_i^2}} \label{4.8}
\end{eqnarray}
where $V$ represents the volume measured in the restframe of the gas.
The integral factorizes using spherical coordinates in momentum space and performing the angle integration gives
\begin{equation}
Z_C = \frac{V^N}{N!}
\left[
4\pi m \int_0^\infty \! d\rho \, \frac{\rho^2}{\sqrt{\rho^2 + m^2}}
\,\mbox{e}^{-\beta \sqrt{\rho^2 + m^2}}
\right]^N
\label{4.9}
\end{equation}
with $\rho=\sqrt{{\it {\bf p}}^2}$. Note, the particle index becomes superfluous considering
a monatomic gas of particles with mass $m_i=m$.
The integral in this expression can be solved applying the transformation
$\rho=m\mbox{sinh}\xi$ (see e.g. \cite{GR})
\begin{eqnarray}
\int_0^\infty \! d\rho \, \frac{\rho^2}{\sqrt{\rho^2+m^2}}
\,\mbox{e}^{-\beta \sqrt{\rho^2+m^2}}
& = &
\int_0^\infty \! d\xi \, m^2 \mbox{sinh}^2\xi
\mbox{e}^{-\beta m \mbox{cosh}\xi} \nonumber \\
& = &
\frac{m^2}{\beta m} K_1(\beta m) \label{4.10}
\end{eqnarray}
with $K_1$ being the modified Bessel function of first order. \\
Finally, the canonical partition integral within this semi-covariant approach is given
by
\begin{equation}
Z_C = \frac{V^N}{N!}
\left[
\frac{4\pi m^3}{\beta m} K_1(\beta m)
\right]^N.
\label{4.11}
\end{equation}
Now we are able to list the analytical results of 3 different approaches:
\begin{equation}
\label{4.12}
Z_C =
  \begin{cases}
    \frac{V^N}{N!}
    \left[
    \frac{4\pi m^3}{\beta m} K_1(\beta m)
    \right]^N  & \mbox{ ; semi-covariant} \\
    \frac{V^N}{N!}
    \left[
    \frac{4\pi m^3}{\beta m} K_2(\beta m)
    \right]^N  & \mbox{ ; J\"uttner approach} \\
    \frac{V^N}{N!}
    \left[ \left( \frac{2\pi m^2}{\beta m}\right)^{3/2} \right]^N & \mbox{ ; non-relativistic
    (see e.g. \cite{Huang}). } \\
  \end{cases}
\end{equation}

Comparing (\ref{4.11}) with the result of J\"uttner we realize that the only difference
is the order of the modified Bessel function occurring in $Z_C$: J\"uttner's approach contains
$K_2(\beta m)$ instead of $K_1(\beta m)$.\\
We note further that this result is naturally quite similar to the one derived by Horwitz et al. (compare eq. (3.25) in \cite{Horwitz81}),
especially regarding the functional dependence on $ \beta $. But we stress that the phase space reduction regarding
the time constraints in our semi-covariant approach is already more satisfying and as a consequence, in contradiction to \cite{Horwitz81}, no commen time interval is remaining in the final result when
applying the formalism of constrained Hamiltonian dynamics.
%%%%%%%%%%%%%%%%%%%%%%%%%%%%%%%%%%%%%%%%%%%%%%%%%%%%%%%%%%%%%%%%%%%%%%
\subsection{Full covariant Approach}
Evaluating the partition integral within the full covariant approach we apply the covariant
time constraints (\ref{2.3}) as specified in section 2.1 for a system of non-interacting particles and equation (\ref{3.11}) reads
\begin{equation}
Z_C = \frac{1}{N!}\int \prod_{i=1}^N d^4\!q_i d^4\!p_i
\,\Theta(p_i^0) \delta\left(\frac{1}{2m_i}(p_i^2-m_i^2)\right)
\delta\left(\frac{1}{m_i}q_i^{\mu}p_{i{\mu}}-\tau\right)
\,\mbox{e}^{-\beta U_{\mu}P^{\mu}}.
\label{4.13}
\end{equation}
Using the properties (\ref{4.d1}) and (\ref{4.d2}) of the
$\delta$-function we get
\begin{equation}
Z_C = \frac{1}{N!}\int \prod_{i=1}^N d^4\!q_i d^4\!p_i
\, \frac{m_i^2}{{p_i^0}^2}
\delta\left(p_i^0-\sqrt{{\it {\bf p}}_i^2+m_i^2}\right)
\delta\left(q_i^0-\frac{1}{p_i^0}({{\it {\bf q}}}{{\it {\bf p}}} - m\tau)\right)
\,\mbox{e}^{-\beta U_{\mu}P^{\mu}} .
\label{4.14}
\end{equation}

Until yet we have kept strictly the covariant fashion during the calculation.
For the further concrete calculation the fixation of a defined frame of reference is demanded.
Hence, for simplicity, we choose the restframe of the gas to determine the invariant energy
$E$ as already applied within the semi-covariant approach (see eq. (\ref{4.7})) and get
integrating the 0-components according the $\delta$-functions
\begin{eqnarray}
Z_C
& = &
\frac{1}{N!} \int \prod_{i=1}^N d^3\!q_i d^3\!p_i
\,\frac{m_i^2}{({\it {\bf p}}_i^2 + m_i^2)}
\,\mbox{e}^{-\beta \sqrt{{\it {\bf p}}_i^2 + m_i^2}} \nonumber \\
& = &
\frac{V^N}{N!} \int \prod_{i=1}^N d^3\!p_i 
\,\frac{m_i^2}{({\it {\bf p}}_i^2 + m_i^2)}
\,\mbox{e}^{-\beta \sqrt{{\it {\bf p}}_i^2 + m_i^2}}.
\label{4.16}
\end{eqnarray}
The integral factorizes using spherical coordinates in momentum space and performing the angle integration gives
\begin{equation}
Z_C = \frac{V^N}{N!}
\left[
4\pi m^2 \int_0^\infty \!d\rho \,\frac{\rho^2}{\rho^2+m^2}
\,\mbox{e}^{-\beta \sqrt{\rho^2 + m^2}}
\right]^N
\label{4.17}
\end{equation}
with $\rho=\sqrt{{\it {\bf p}}^2}$. Note, as in equation (\ref{4.9}), $m_i\!=\!m$ has been used again.

Unfortunately, no analytical solution of the integral in equation (\ref{4.17})
is available. Hence, numerical solutions have been worked out for  various monatomic gases.

In order to compare this numerical results obtained by the full covariant approach with
the results of other approaches as listed in equation (\ref{4.12}) we write the canonical
partition integral of the monatomic gas in a generic representation:
\begin{equation}
Z_C = \frac{V^N}{N!}\,(\mbox{e}^{-\beta m}Y)^N.
\label{4.18}
\end{equation}

The 'relativistic renormalization factor' $\mbox{e}^{-\beta m}$ has been extracted
explicitly in (\ref{4.18}) for two reasons:\\
1) This factor does not occur in the non-relativistic partition integral. \\
2) The numerical integration is simplified if this factor is extracted
explicitly.

The non-analytical integral occurring in the full covariant approach (see eq. (\ref{4.17}))
has been determined numerically using the Gauss-Laguerre method of $15^{th}$
order. The results of the two other relativistic approaches
(i.e. the J\"uttner approach and the semi-covariant approach) have been
determined using the analytical expressions (i.e. evaluating the related
modified Bessel functions $K_1$ and $K_2$) as well as using the same numerical
integration method in order to justify the accuracy of the applied numerical method.

Figure 1 summarizes the results of calculations for different monatomic perfect gases nameley, hydrogen, helium, neon and argon.
The comparison is focused on the quantity $Y$ which determines the canonical partition integral via equation (\ref{4.18}).
As expected, the qualitative behavior is the same for all gases.

\begin{figure}
\begin{center}
\fbox{\scalebox{0.65}{\includegraphics{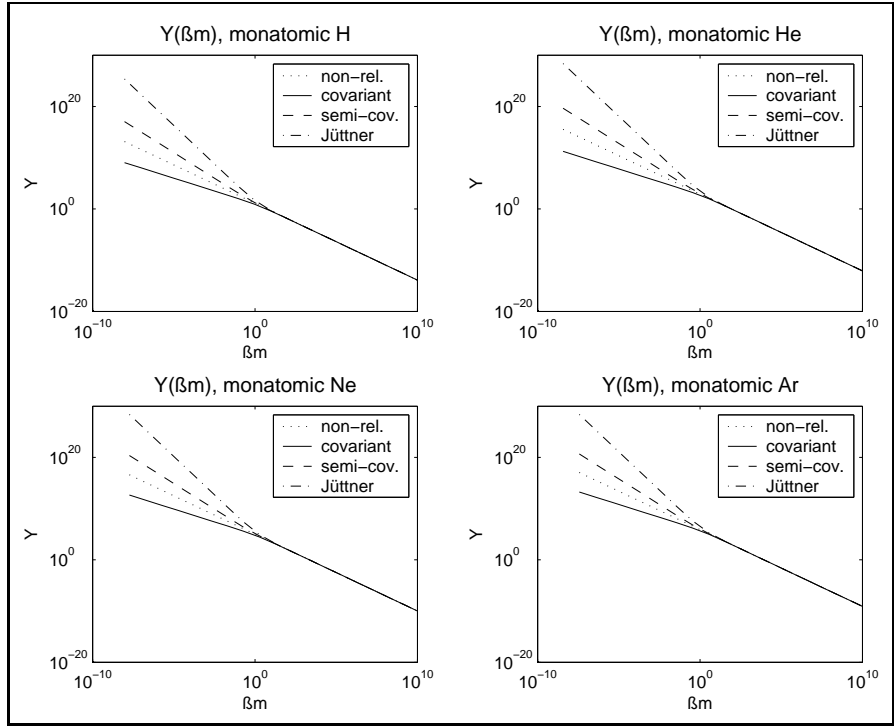}}}
\end{center}
\caption{Quantity $Y$ as a function of ${\beta}m$ for various approaches and different monatomic perfect gases.}
\end{figure}

At low temperatures (non-relativistic limit) all approaches give the same
result as it should be. Even at higher temperatures
($T \leq 10^{12}K$)\footnote{E.g. $T = 10^{12} K$ corresponds roughly to $\beta m \simeq 10$ in case of hydrogen.}
no difference is visible among the various calculations.
At extremely high temperatures ($T \geq 10^{12}K$) relativistic effects
are dominating the behavior.
Compared to the non-relativistic result the non-covariant J\"uttner
approach is showing
a strong enhancement of the quantity $Y$ in this regime. The semi-covariant
approach reduces this enhancement and the full covariant approach turns the
enhancement to a reduction even.
Therefore, we conclude that a controlled manifest covariant
reduction of the extended phase space reduces the value of the quantity $Y$
and hence of the canonical partition integral.

In order to give an overview on the related thermodynamics of the monatomic perfect gas we finally list in dependence of the quantity $Y$ the free energy
\begin{equation}
F
=
- kT \mbox{ln} Z_C
=
- NkT\left( \mbox{ln}\frac{V}{N}+ \mbox{ln}Y - \beta m +1\right) ,
\label{4.20}
\end{equation}
the entropy
\begin{eqnarray}
S
& = &  
- \left( \frac{\partial F}{\partial T}\right)_V
=
k \mbox{ln} Z_C  + kT \frac{\partial \mbox{ln}Z_C}{\partial T} \nonumber \\
& = &
Nk \left( \mbox{ln}\frac{V}{N}+ \mbox{ln}Y
-\beta \frac{\partial \mbox{ln}Y}{\partial \beta} +1 \right)
\label{4.21}
\end{eqnarray}
the pressure
\begin{equation}
P
=
- \left( \frac{\partial F}{\partial V}\right)_T
=
kT \frac{\partial \mbox{ln}Z_C}{\partial V}
=
\frac{NkT}{V}, \label{4.22}
\end{equation}
the average energy
\begin{equation}
\left\langle E\right\rangle
= N \left(m - \frac{\partial \mbox{ln}Y}{\partial \beta}\right) ,
\label{4.23}
\end{equation}
and the specific heat
\begin{equation}
c_V
=
\left( \frac{\partial \left\langle E\right\rangle }{\partial T}\right)_V
=
\frac{N\beta}{T} \frac{\partial^2 \mbox{ln}Y}{\partial \beta^2}.
\label{4.24}
\end{equation}

Note, since the pressure turns out to be independent of $Y$ all approaches
give the same equation of state,
namely the one of the non-relativistic perfect gas:
\begin{equation}
PV = NkT.
\label{4.25}
\end{equation}

Since the main differences among the various approaches are showing up at very high temperatures
it is very usefull to consider explicitly the
\\  
%%%%%%%%%%%%%%%%%%%%%%%%%%%%%%%%%%%%%%%%%%%%%%%%%%%%%%%%%%%%%%%%%%%%%%
\subsection{Ultra-relativistic Limit}

Since in the ultra-relativistic limit the rest mass is neglected in respect to the momentum
and in respect to the energy it can be evaluated using
$ \sqrt{{\it {\bf p}}_i^2+m_i^2} \longrightarrow |{\it {\bf p}}_i| $ when deriving the integrals in eqs.
(\ref{4.17}) and (\ref{4.9}).
Hence, it is even possible to derive analytical results in case of the full covariant approach
since the partition integral simplifies drastically and we get in this limit ($\rho=|{\it {\bf p}}|$) for the various approaches:
\begin{equation}
\label{4.26}
Z_C^{(ur)} =
  \begin{cases}
    \frac{V^N}{N!} \left[ 4\pi m^2 \int_0^\infty \! d\rho
      \,\mbox{e}^{-\beta \rho} \right]^N
   & \mbox{ ; full covariant} \\
    \frac{V^N}{N!} \left[ 4\pi m \int_0^\infty \! d\rho \rho
      \,\mbox{e}^{-\beta \rho} \right]^N
   & \mbox{ ; semi-covariant} \\
    \frac{V^N}{N!} \left[ 4\pi  \int_0^\infty \! d\rho \rho^2
      \,\mbox{e}^{-\beta \rho} \right]^N
   & \mbox{ ; non-covariant (J\"uttner)} \\
  \end{cases}
\end{equation}
The integrals in (\ref{4.26}) are easily determined with help of the $\Gamma$-function according
\begin{equation}
\label{4.28}
\int_0^\infty\! d\zeta \zeta^n \,\mbox{e}^{-a \zeta}
= \frac{\Gamma(n+1)}{a^{n+1}} .
\end{equation}
Table 1 summarizes these results as well as some derived thermodynamical quantities
in case of the ultra-relativistic limit.
Progressing from the non-covariant approach towards the full covariant approach via the
semi-covariant approach at a fixed $\beta$ the canonical partition integral is decreasing
as already mentioned when discussing figure 1. This behavior as well as the derivatives
in respect to $\beta$ are fixing the differences in the thermodynamical quantities,
e.g. the entropy is reducing when progressing towards the full covariant approach.
Especially, we want to focus on the average energy:\\
While J\"uttners non-covariant calculations at the ultra-relativistic limit lead to $3NkT$
the semi-covariant approach (as well as the method used in \cite{Horwitz81}) leads to $2NkT$
and the full covariant treatment gives $NkT$.
Thus, a controlled manifest covariant reduction of the extended phase space by $ N $ on-shell constraints reduces the average energy
by a factor $ NkT $ and the final reduction by $ N $ time fixations
by another factor $ NkT $.
Consequently, the specific heat is showing similar differences.
As already mentioned by Horowitz et al. \cite{Horwitz81} no empirical evidence distinguishing between these results is available at the present time.\\
The following should be noted:\\
Usually, the non-covariant J{\"u}ttner approach in the 
ultra-relativistic limit is regarded to represent the classical photon gas. This approximate interpretation is not possible
within our approach because the applied classical constrained Hamiltonian dynamics of massive particles can not be used in the ultra-relativistic limit to approximate massless classical particle dynamics.
This fact is showing up in the results and can be recognized inspecting $Z_C^{(ur)}$ (see table 1):
In contradiction to the J{\"u}ttner approach $Z_C^{(ur)}$ remains mass dependant in the ultra-relativistic limit in case of the semi-covariant and the full covariant approach.
\\
\small
\begin{center}
\begin{table}
\begin{tabular}{| l || c | c | c |}
\hline
& & & \\
thermodynamical & full covariant & semi-covariant & non-covariant \\
quantity & & & (J\"uttner approach) \\
& & & \\
\hline
\hline
& & & \\
$Z_C^{(ur)}$ &
$\frac{V^N}{N!} \left[ \frac{4\pi m^3}{(\beta m)} \right]^N $ &
$\frac{V^N}{N!} \left[ \frac{4\pi m^3}{(\beta m)^2} \right]^N $ &
$\frac{V^N}{N!} \left[ \frac{8\pi m^3}{(\beta m)^3} \right]^N $ \\
& & & \\
\hline
& & & \\
$ F = -kT\mbox{ln}Z_C $ &
$ -NkT[ \mbox{ln}\frac{V}{N}+\mbox{ln}(4 \pi m^3) $ &
$ -NkT[ \mbox{ln}\frac{V}{N}+\mbox{ln}(4 \pi m^3) $ &
$ -NkT[ \mbox{ln}\frac{V}{N}+\mbox{ln}(8 \pi m^3) $ \\
$  $ &
$ -\mbox{ln}(\beta m) + 1 ] $ &
$ -2 \mbox{ln}(\beta m) + 1 ] $ &
$ -3 \mbox{ln}(\beta m) + 1 ] $ \\
& & & \\
\hline
& & & \\
$ S = k\mbox{ln}Z_C $ &
$ Nk[ \mbox{ln}\frac{V}{N}+\mbox{ln}(4 \pi m^3) $ &
$ Nk[ \mbox{ln}\frac{V}{N}+\mbox{ln}(4 \pi m^3) $ &
$ Nk[ \mbox{ln}\frac{V}{N}+\mbox{ln}(8 \pi m^3) $ \\
$ -\frac{1}{T}{\partial \mbox{ln}Z_C}/{\partial \beta} $ &
$ -\mbox{ln}(\beta m)   + 2 ] $ &
$ -2 \mbox{ln}(\beta m) + 3 ] $ &
$ -3 \mbox{ln}(\beta m) + 4 ] $ \\
& & & \\
\hline
& & & \\
$ \left\langle E\right\rangle = - {\partial \mbox{ln}Z_C}/{\partial \beta}$ &
$ NkT  $ &
$ 2NkT $ &
$ 3NkT $ \\
& & & \\
\hline
& & & \\
$ c_V = {\partial\!\left\langle E\right\rangle}/{\partial T} $ &
$ Nk  $ &
$ 2Nk $ &
$ 3Nk $ \\
& & & \\
\hline
& & & \\
$ P = -{\partial F}/{\partial V} $&
$ NkT/V $ &
$ NkT/V $ &
$ NkT/V $ \\
& & & \\
\hline
\end{tabular}
\caption{Thermodynamical quantities for the perfect gas in the ultra-relativistic limit}
\end{table}
\end{center}
\vspace{2mm}
\normalsize
%%%%%%%%%%%%%%%%%%%%%%%%%%%%%%%%%%%%%%%%%%%%%%%%%%%%%%%%%%%%%%%%%%%%%%
\subsection{Non-relativistic Limit}

Finally, we examine the non-relativistic limit of our approach. Since the covariant time fixations (\ref{2.3}) reduce to the
simple time fixations (\ref{2.2}) in this limit the canonical partition
integral within both approaches, semi-covariant and full covariant,
coincide according eq. (\ref{2.11}) to be
\begin{equation}
\label{4.30}
Z_C^{(nr)} = \frac{1}{N!}\int \prod_{i=1}^N d^4\!q_i d^4\!p_i
\,\mbox{e}^{-\beta E}
\delta\left( \varepsilon_i - \frac{{\it {\bf p}}_i^2}{2m_i}\right)  \delta(q_i^0-\tau).
\end{equation}
Special care has to be taken regarding the invariant energy $E$.
In order to get a fair comparison in the non-relativistic limit
the rest masses of the particles have to be subtracted from their energies getting in the comoving frame of reference
\begin{equation}
\label{4.31}
E = \sum_i^N(p_i^0-m_i).
\end{equation}
Using further $ p_i^0 = \varepsilon_i+m_i $ equation (\ref{4.30}) simplifies to
\begin{equation}
\label{4.32}
Z_C^{(nr)}
=
\frac{V^N}{N!}\int \prod_{i=1}^N d^3\!p_i
\,\mbox{e}^{-\beta \frac{{\it {\bf p}}_i^2}{2m_i}}
=
\frac{V^N}{N!}\left[ \left( \frac{2 \pi m^2}{\beta m}\right)^{3/2}\right]^N ,
\end{equation}
i.e. the non-relativistic canonical partition integral.

Note, the same result can be achieved starting with the analytical result as obtained by the semi-covariant approach. Supplementing
eq. (\ref{4.11}) by a factor $\mbox{e}^{\beta Nm}$ which covers the effect of the reduction of the energies by the rest masses
(compare eq. (\ref{4.31})) equation (\ref{4.11}) transforms to
\begin{equation}
\label{4.33}
Z_C
=
\frac{V^N}{N!} \mbox{e}^{\beta Nm}
\left[ \frac{4\pi m ^3}{\beta m} K_1(\beta m)\right]^N.
\end{equation}
In the non-relativistic limit ($m\!\gg\!kT$) the asymptotic form
of the modified Bessel function ($\beta m \rightarrow \infty $)
\begin{equation}
\label{4.34}
K_n(\beta m) \simeq \sqrt{\frac{\pi}{2 \beta m}} \mbox{e}^{-\beta m}
\left(1+\frac{4n^2-1}{8\beta m}+...\right)
\end{equation}
can be applied.
Since $K_1$ and $K_2$ coincide in the leading order the semi-covariant
approach and the non-covariant J\"uttner approach provide the same result in the non-relativistic limit namely, the non-relativistic
canonical partition integral (\ref{4.32}) within this treatment.
%%%%%%%%%%%%%%%%%%%%%%%%%%%%%%%%%%%%%%%%%%%%%%%%%%%%%%%%%%%%%%%%%%%%%%
%%%%%%%%%%%%%%%%%%%%%%%%%%%%%%%%%%%%%%%%%%%%%%%%%%%%%%%%%%%%%%%%%%%%%%
\newpage
\section{Summary and Conclusion}

A covariant equilibrium statistical mechanics has been formulated on the
base of constrained Hamiltonian dynamics.
The usage of this formalism to describe the micro-dynamics of the ensembles
guarantees the manifest covariance of the developed approach.

Unfortunately, concrete calculations are not easy performed within this approach.
As a simple application the canonical partition integral of the
monatomic perfect gas has been evaluated numerically.
Relativistic effects at very high temperatures have been observed: \\
The comparison of the obtained results with the results
of the non-relativistic theory show a decreasing partition integral
at very high temperatures.
In contradiction to this findings other relativistic approaches
like the non-covariant calculations of J\"uttner are showing an increased partition
integral at these temperatures.
It has been demonstrated that the full covariant treatment of the phase space
is responsible for this significant difference.
Considering the ultra-relativistic limit the same behavior is visible on a
full analytical base. In the non-relativistic limit the results coincide with the results obtained within the non-relativistic theory.

As in the non-relativistic theory the link to thermodynamics has been performed via the free energy.
All thermodynamical potentials
(like the internal energy, the free energy, ...) as well as the
thermodynamical state variables (like temperature, entropy, pressure, ...)
have been defined to be Lorentz scalars. This was achieved introducing the invariant internal energy $E\!=\!U_{\mu}P^{\mu}$
in the canonical distribution function.
Regarding the ultra-relativistic limit
the entropy, the average energy and the specific heat are
decreased in comparison to other approaches in this limit.

Unfortunately, the temperature regime which shows significant differences
among the various approaches ($T \geq 10^{12}K$) can not be reached
in experiments but may exist in cosmological events.
Hence, it may be interesting to use the presented special relativistic approach
as a base to develop a general relativistic equilibrium statistical mechanics
being manifest covariant from first principle.
\\
\vspace{2cm}
\\
%%%%%%%%%%%%%%%%%%%%%%%%%%%%%%%%%%%%%%%%%%%%%%%%%%%%%%%%%%%%%%%%%%%%
%%%%%%%%%%%%%%%%%%%%%%%%%%%%%%%%%%%%%%%%%%%%%%%%%%%%%%%%%%%%%%%%%%%%%%
{\large \bf Acknowledgement}\\
\\
The author is pleased to thank S. Ulrych for various fruitful discussions.
 
%%%%%%%%%%%%%%%%%%%%%%%%%%%%%%%%%%%%%%%%%%%%%%%%%%%%%%%%%%%%%%%%%%%%%%
%%%%%%%%%%%%%%%%%%%%%%%%%%%%%%%%%%%%%%%%%%%%%%%%%%%%%%%%%%%%%%%%%%%%%%

%%%%%%%%%%%%%%%%%%%%%%%%%%%%%%%%%%%%%%%%%%%%%%%%%%%%%%%%%%%%%%%%%%%%%

\end{document}